\documentclass[aps,prb]{revtex4}
\usepackage{graphicx}
\usepackage{amsmath}
\usepackage{amsfonts}
\usepackage[dvips]{color}
\makeatletter

\newcommand{\Rmnum}[1]{\expandafter\@slowromancap\romannumeral #1@}
\makeatother

\begin{document}

\title{Electron Tunneling in Monolayer and Bilayer Graphene}
\author{Di Wu, Weiqiang Chen and Fu-Chun Zhang}
\affiliation{Department of Physics, and Center for Theoretical and Computational
  Physics, The University of Hong Kong, Pokfulam Road, Hong Kong, China}
\date{\today}

\begin{abstract}

Electron's tunneling through potential barrier in monolayer and bilayer
graphene lattices is investigated by using full tight-binding model. Emphasis is
placed on the resonance tunneling feature and inter-valley scattering
probability. It is shown that normal incidence transmission probabilities for
monolayer and bilayer graphene exhibit different properties. Our calculation
indicates that valleytronics in graphene systems may be detected, generated and
controlled by changing the structure parameters of the external electric potential.

\end{abstract}

% \pacs{}

\maketitle

\section{Introduction}

Recently, great interest has been aroused in research on the physical properties
of graphene, a one-atom-thick allotrope of carbon, due to its successful
fabrication in 2004\cite{Novoselov-2004}. Monolayer graphene is a truly
two-dimensional material, with unusual electronic excitations described in terms
of Dirac fermions that move in a curved space\cite{Wallace-1947}. The electrons
in graphene seem to be almost insensitive to disorder and electron-electron
interactions and have very long mean free path\cite{Pereira-2008}. Hence,
graphene's transport properties are rather different from what is found in usual
metals and semiconductors. Interestingly enough, these properties can be easily
modified with the application of electric and magnetic fields, the addition of
layers, and by controlling its geometry and chemical doping\cite{Geim-2007}. Apart from
the interesting fundamental physics in this new system, graphene is attracting
attention as a promising new material for electronic applications. For a review
concerning the history, fabrication, fundamental properties, and future
applications of graphene, we refer to the recent article\cite{Neto-2008}.

The low energy charge carriers in graphene are described by a massless Dirac
equation and have a linear energy dispersion which is isotropic near the Dirac
points where the valence and conduction bands meet each
other\cite{Novoselov-2005}. Such characteristics offer exciting opportunities
for the occurrence of new tunneling phenomena and the development of high
quality devices. Therefore, it may be valuable to investigate the electronic
transport properties of graphene. In this regards, much of the phenomena
associated with tunneling in graphene systems has been theoretically
studied\cite{Katsnelson-2006,Pereira-2006,Cheianov-2006,Bai-2007,Stander-2008}. It is
interesting that, owing to the chiral nature of the quasiparticles, the
propagation of charge carriers in monolayer graphene mimics the tunneling of
massless fermions. This relativistic effect provides us an experiment test for
the Klein paradox\cite{Klein-1929,Katsnelson-2006}, which predicts that electron
can pass through a high potential barrier to approach the perfect
transmission. In contrast, for conventional non-relativistic particles, the
transmission probability exponentially decays with the increasing of the barrier
height. Besides this relativistic transport feature, other promising tunneling
properties of graphene systems are the ability to tune the carrier density
through a gate voltage\cite{Novoselov-2004}, the absence of back
scattering\cite{Ando-1998}, and the fact that graphene exhibits both spin and
valley degrees of freedom\cite{Di-2007}, which might be harnessed in envisaged
spintronic\cite{Kane-2005,Cho-2007} or valleytronic devices\cite{Rycerz-2007}.

However, in these previous studies, considering the carriers as massless
fermions remains an approximation, and we expect a deviation from a linear
dispersion for high energies of the Dirac cone\cite{Plochocka-2008}. To this
end, it is both important and interesting to study the transport properties in
graphene system without this approximation. In this work, we investigate the
tunneling properties of both monolayer and bilayer graphene lattice by using the
full tight-binding model. This paper is organized as follows. In
Sec.~\ref{sec:model}, we introduce the model of the system that we are
considering and give explicit expressions for the wave functions in different
regions through tunneling process. We also provide some details on how
to compute the transmission amplitudes. In Sec.~\ref{sec:result}, we present the
results for transmission probabilities in different barrier setups and the
discussions. A brief summary and conclusion of the paper can be found in
Sec.~\ref{sec:summary}.

\section{Model}
\label{sec:model}

\begin{figure}[htbp]

\resizebox{80mm}{!}{\includegraphics[]{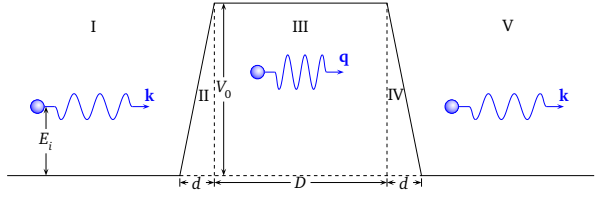}}

\caption[]{\label{fig:potential} Schematic of the external potential profile in
  graphene lattice.  The energy of incident electron is $E_i$ and the potential
  height is $V_0$. There are five different regions in this tunneling process
  which are marked by roman numbers.}

\end{figure}

We consider two kinds of lattice structure, each consisting respectively of
monolayer graphene or bilayer graphene.  A schematic picture of the system is
shown in Fig.~\ref{fig:potential}. The basic structure involves a graphene sheet
and a one dimensional trapezoid shape potential $V(\mathbf{r}) = V(x)$, which is
$y-$independent.  In the following we will introduce the tight-binding model of
graphene with this external electric potential.

\subsection{Tunneling in monolayer graphene lattice}

\begin{figure}[htbp]

 $\begin{array}{cc}
    \resizebox{40mm}{!}{\includegraphics[]{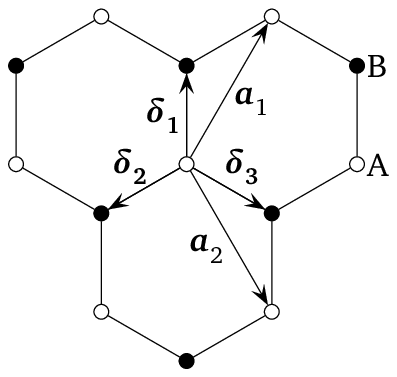}} &
    \resizebox{40mm}{!}{\includegraphics[]{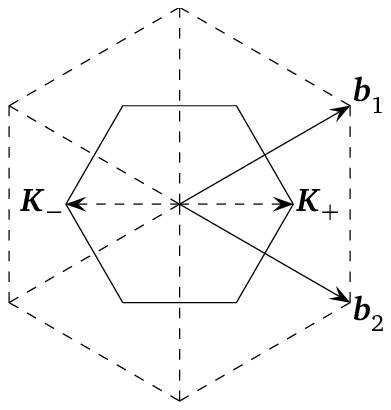}}
  \end{array}$\\
  \resizebox{80mm}{!}{\includegraphics[]{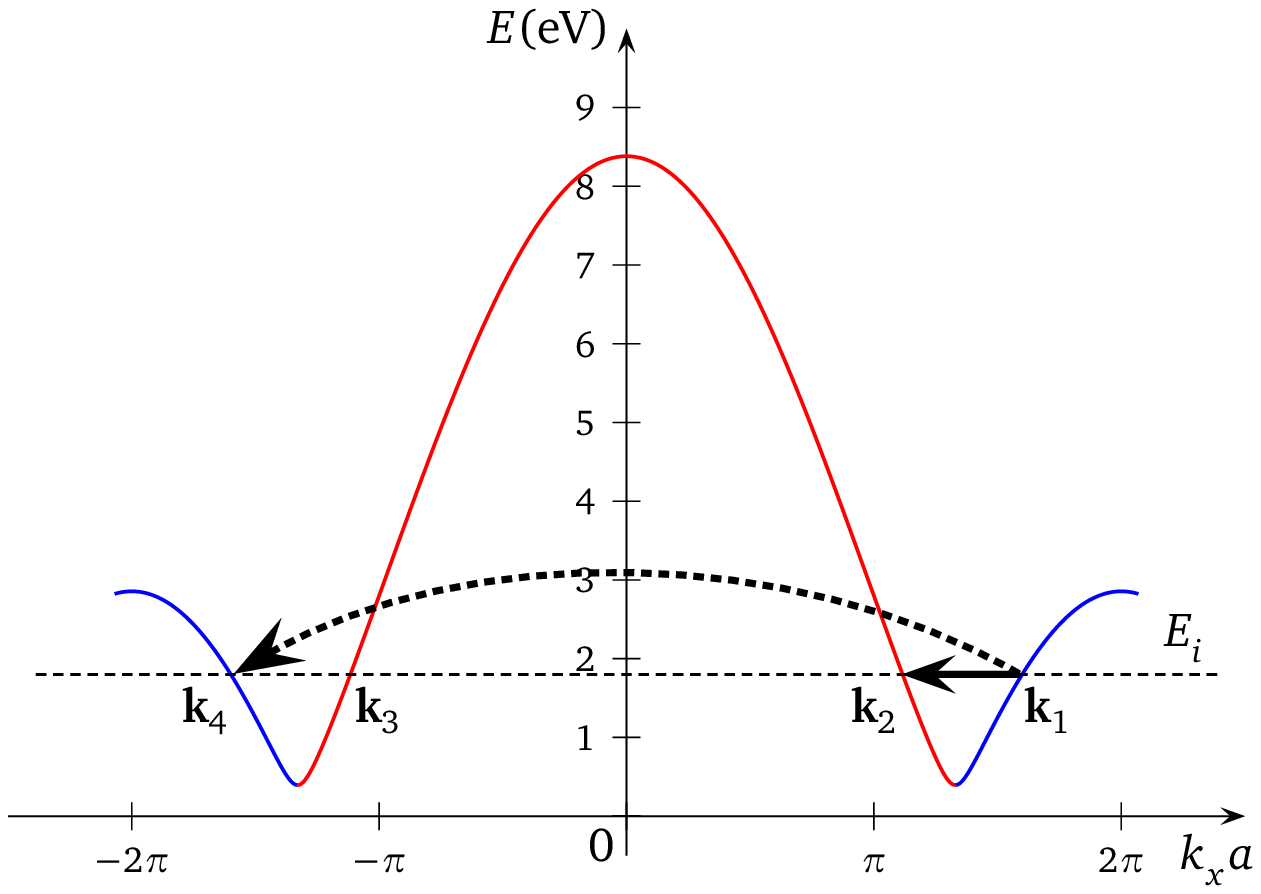}}

\caption[]{\label{fig:monolayer} Schematics of the lattice structure and the
  energy spectrum of monolayer graphene.  Top left: Lattice structure of
  monolayer graphene, made out of two interpenetrating triangular
  lattice($\mathbf{a}_1$ and $\mathbf{a}_2$ are the lattice unit vectors, and
  $\boldsymbol{\delta}_i$, $i = 1, 2, 3$ are the nearest neighbor vectors). Top
  right: The solid hexagon is a Brillouin zone. Dirac cones sit at the
  $\mathbf{K}_{+}$ and $\mathbf{K}_{-}$ points. The reciprocal lattice vectors
  are $\mathbf{b}_1$ and $\mathbf{b}_2$. Bottom: The energy spectrum of
  electrons at a finite $k_y$. There are four degenerate states for a given
  $k_y$ when incident energy $E_i$ is small. Two of these states($\mathbf{k}_1$
  and $\mathbf{k}_4$) have pseudo-spin $+1$, the other two states($\mathbf{k}_2$
  and $\mathbf{k}_3$) have pseudo-spin $-1$. The solid arrow is the intra-valley
  scattering process(flip pseudo-spin) while the dashed arrow is the
  inter-valley scattering process(does not flip pseudo-spin).}

\end{figure}

The honeycomb lattice of monolayer graphene can be described in terms of two
triangular sublattices, A and B(see Fig.~\ref{fig:monolayer}). A unit cell
contains two atoms, one of type A and one of type B. The lattice vectors can be
written as
$\mathbf{a}_1 = a(1/2, \sqrt{3}/2), \mathbf{a}_2 = a(-1/2,\sqrt{3}/2)$,
where $a \approx 2.46\text{\AA}$ is the lattice
constant\cite{Castro-2007}. The reciprocal lattice vectors are given by
$\mathbf{b}_1 = \frac{2\pi}{a}(1, 1/\sqrt{3})$ and
$\mathbf{b}_2 = \frac{2\pi}{a}(1, -1/\sqrt{3})$.

In monolayer graphene, an atom of type A is connected to its nearest neighbors on B
sites by three vectors $\boldsymbol{\delta}_i$.  The nearest hopping
tight-binding Hamiltonian describing this system has the form\cite{Neto-2008}
\begin{align}
\label{eq:1} H_0 = -t \sum_{\langle i,j\rangle,\sigma} \left(
a^{\dagger}_{\sigma,i}b_{\sigma,j} + \text{h.c.} \right),
\end{align}
where $t(\approx 2.8\text{eV})$ is the nearest neighbor hopping
energy, $a_{\sigma,i}$ and $b_{\sigma,j}$ are the annihilation operators of
electrons with spin $\sigma(\sigma = \uparrow,\downarrow)$ on A and B
sublattices, respectively. In momentum representation, the Hamiltonian reads
$H_0 = \sum_{\mathbf{k},\sigma} \psi^{\dagger}_{\sigma} \left( \mathbf{k}
\right) \mathcal{H}_0 \psi_{\sigma} \left( \mathbf{k} \right)$, where
\begin{align}
\label{eq:2}
\mathcal{H}_0 =
\begin{pmatrix}
0 & \phi^{*} \left( \mathbf{k} \right) \\
\phi \left( \mathbf{k} \right) & 0
\end{pmatrix}
\end{align}
with
$\phi \left( \mathbf{k} \right) = -t \sum_{\boldsymbol{\delta}_i} e^{-i
  \mathbf{k}\boldsymbol{\delta}_i} \equiv - \epsilon \left( \mathbf{k}
\right)e^{i\varphi \left( \mathbf{k} \right)} $.
This Hamiltonian acts on a spinor representing the wave function on two
sublattices:
$\psi_{\sigma} = \left( \psi_{\sigma\text{A}}, \psi_{\sigma\text{B}} \right)^T$.
Accordingly, quasiparticle's excitation spectrum has two branches(bands) with
the dispersion
$E^{\pm} \left( \mathbf{k} \right) = \pm \epsilon \left( \mathbf{k} \right)$
and\cite{Neto-2008}
\begin{align}
\label{eq:3}
\epsilon \left( \mathbf{k} \right) = t \sqrt{1 + 4 \cos \frac{k_xa}{2} \cos
  \frac{\sqrt{3}k_ya}{2} + 4\cos^2 \frac{k_xa}{2}}.
\end{align}
The upper band($E^{+}$) and the lower band($E^{-}$) meet at six corners of the
first Brillouin zone. Only two out of the six are independent, which we choose
to be $\mathbf{K}_{\pm}=\pm \frac{2\pi}{a}(2/3,0)$ as shown in
Fig.~\ref{fig:monolayer}. With an expansion around $\mathbf{K}_{\pm}$, one can
find that the Hamiltonian (2) reduces to a massless Dirac Hamiltonian with
linear dispersion $E^{\pm}\left( \mathbf{k} \right) = \pm \hbar
v_F|\mathbf{k}|$. It is important to note that in monolayer graphene the
pseudo-spin is defined as eigenvalue of the Pauli matrix $\sigma_1$.

In tunneling problem, we consider the barrier of which the geometry is shown in
Fig.~\ref{fig:potential}. The dynamics is governed by the Schr\"{o}dinger
equation for an incident particle of energy $E$,
\begin{align}
\label{eq:4}
\left[ H_0 + V \left( x \right)\right] \psi = E \psi.
\end{align}
We solve eqn.~\eqref{eq:4} by using the standard method. At first, we assume the
incident electron wave propagates along the $x$ axis with a given energy $E_i$
and wave vector $k_i$. Because $V$ has no $y$ dependence, the wave vector along
$y$ direction is conserved, so the wave vector along the $y$-direction in all the
regions remains to be $k_{iy}$ . In region \Rmnum{1} and \Rmnum{5}, the
eigenstates with $\epsilon(\mathbf{k}) = E_i$ and $k_y = k_{iy}$ are fourfold
degenerate as shown in Fig.~\ref{fig:monolayer}, while the corresponding
eigenfunctions are
\begin{align}
\label{eq:5}
\psi \left( \mathbf{k}_l \right) = \frac{1}{\sqrt{2}}
\begin{pmatrix}
1\\
\phi \left( \mathbf{k}_l \right)/E_i
\end{pmatrix}e^{i \mathbf{k}_l \mathbf{r}}.
\end{align}

Inside the barrier, i.e, in region \Rmnum{3}, the electron's wave vectors should
satisfy $\epsilon \left( \mathbf{q}_l \right) = V_0 - E_i, l = 1, 2, 3, 4$ and
$q_{ly} = k_{iy}$. The eigenfunctions inside the barrier can be written as
\begin{align}
\label{eq:6}
\psi \left( \mathbf{q}_l \right) = \frac{1}{\sqrt{2}}
\begin{pmatrix}
1\\
\phi \left( \mathbf{q}_l \right)/ \left( E_i - V_0 \right)
\end{pmatrix}e^{i \mathbf{q}_l \mathbf{r}}.
\end{align}

After constructing all these wave functions, the next step is to determine
incident and reflected waves through the whole tunneling process which can be
done with the velocities of these states
\begin{align}
\label{eq:7}
& v_x \left( \mathbf{k} \right) = \frac{1}{\hbar} \frac{\mathrm{d} \epsilon
  \left( \mathbf{k} \right)}{\mathrm{d} k_x}\notag\\
= & -\frac{at}{\hbar} \frac{\sin \left( \frac{k^{}_xa}{2} \right) \left[ \cos
    \left( \frac{\sqrt{3}k^{}_ya}{2} \right)+2\cos \left( \frac{k^{}_xa}{2}
    \right) \right]} {\sqrt{1 + 4 \cos \left( \frac{k^{}_xa}{2} \right) \cos
    \left( \frac{\sqrt{3}k^{}_ya}{2} \right) + 4\cos^2 \left( \frac{k^{}_xa}{2}
    \right)}}.
\end{align}
Therefore, for the right movers in Fig.~\ref{fig:potential}, we
require\cite{Nilsson-2007} $v_x \left( \mathbf{k} \right) > 0$; on the contrary,
for the left movers, their velocity $v_x \left( \mathbf{k} \right) < 0$.

Before proceeding to the numerical calculation, it will be instructive to make
some general analyses about the two different scattering processes in this
system as shown in Fig.~\ref{fig:monolayer}. In that figure, the solid arrow
represents intra-valley scattering, while the dashed arrow represents
inter-valley scattering. The latter one has a much larger momentum transfer
($\sim 2\left|\mathbf{K}_{+}\right|$), hence it is usually neglected for low
barrier ($V_0 \ll t$)
tunneling\cite{Katsnelson-2006,Bai-2007,Nilsson-2007,Park-2008}. Furthermore,
these two scattering processes are associated with the two operations on the
pseudo-spin of the quasiparticles: intra-valley scattering flips the
pseudo-spin; while during inter-valley scattering the pseudo-spin is conserved.

We can write down the general solutions in different
regions\cite{Katsnelson-2006,Bai-2007}. In region \Rmnum{1}, we have one
incident wave ($\mathbf{k}_1$ or $\mathbf{k}_3$) and two reflective waves
($\mathbf{k}_2$ and $\mathbf{k}_4$), and the solutions can be expressed as
\begin{align}
\label{eq:8}
\Psi_{\text{\Rmnum{1}}} \left( \mathbf{r} \right) = \psi \left( \mathbf{k}_1
\right) + r_1\psi \left( \mathbf{k}_2 \right) + r_2 \psi \left( \mathbf{k}_4
\right).
\end{align}
Here $r_1$ is the intra-valley reflection amplitude and $r_2$ is the
inter-valley reflection amplitude. In region \Rmnum{3} we have four hole states
with
\begin{align}
\label{eq:9}
\Psi_{\text{\Rmnum{3}}} \left( \mathbf{r} \right) = f_1 \psi\left( \mathbf{q}_1
\right) + f_2 \psi\left( \mathbf{q}_2 \right) + f_3 \psi\left( \mathbf{q}_3
\right) + f_4 \psi\left( \mathbf{q}_4 \right).
\end{align}
Finally, in region \Rmnum{5}, we have two transmitted waves
\begin{align}
\label{eq:10}
\Psi_{\text{\Rmnum{5}}} \left( \mathbf{r} \right) = t_1\psi \left( \mathbf{k}_1
\right) + t_2 \psi \left( \mathbf{k}_3 \right).
\end{align}

Now we need to solve the Schr\"{o}dinger equation in region \Rmnum{2} and
\Rmnum{4} numerically to find the wavefunctions. Upon applying the
continuity of the wave function at the boundaries, one may obtain the
coefficients $r$, $f$, and $t$. After that one can determine the transport
probability by
$T = |t_1|^2 + |t_2|^2 \times v_x \left( \mathbf{k}_3 \right) / v_x \left(
  \mathbf{k}_1 \right)$.
An important advantage of our model is that we can choose the incoming electron
from different valleys($\mathbf{k}_1$ or $\mathbf{k}_3$), which would lead to
different tunneling properties in the results.

\subsection{Tunneling in bilayer graphene lattice}

\begin{figure}[htbp]

 $\begin{array}{cc}
    \resizebox{40mm}{!}{\includegraphics[]{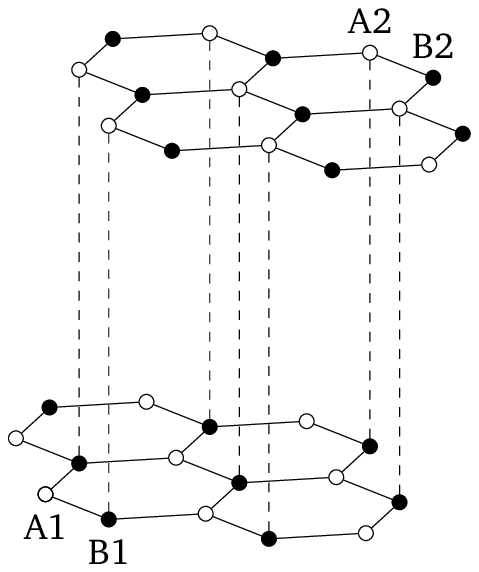}} &
    \resizebox{40mm}{!}{\includegraphics[]{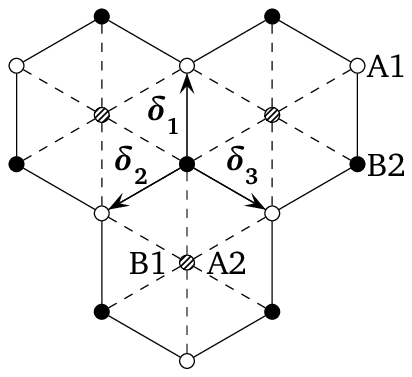}}
  \end{array}$\\
  \resizebox{80mm}{!}{\includegraphics[]{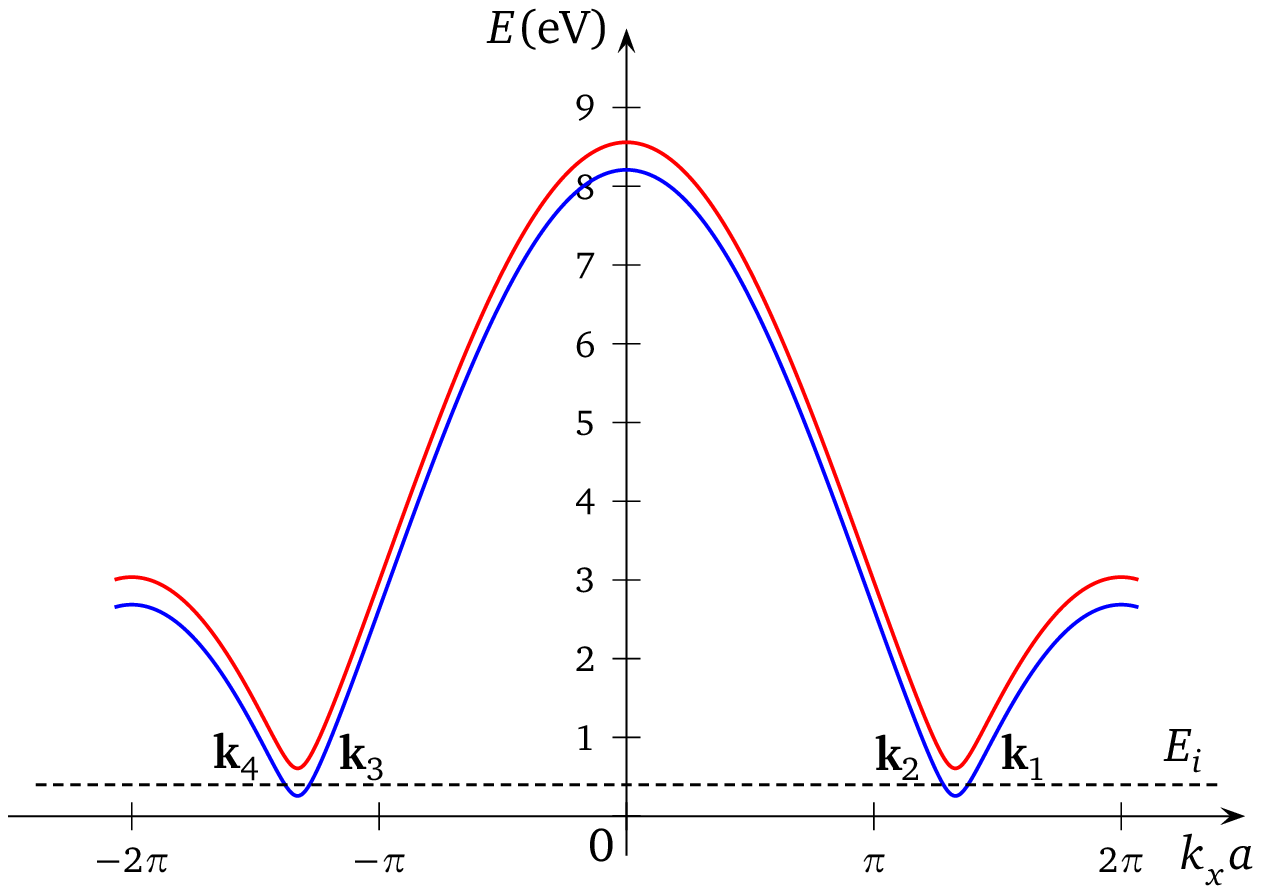}}

\caption[]{\label{fig:bilayer} Schematics of the lattice structure and the
  energy spectrum of bilayer graphene. Top left: Interlayer coupling $t_{\perp}$
  forms 'dimers' from pairs of A2-B1 orbitals(dashed lines), which leading to
  the formation of high energy bands. Top right: Top view of bilayer
  lattice. The only hopping mechanism between two layers is
  A1(open)$\rightleftharpoons$B2(solid) hopping via A2-B1(hashed) dimer
  state. Bottom: Band structure of bilayer graphene. At the Dirac point, the
  spectrum has a gap of $t_{\perp}$. The pseudo-spin of lower band is $+1$, of
  upper band is $-1$.}

\end{figure}

Bilayer graphene consists of two monolayer stacked as in natural graphite(see
Fig.~\ref{fig:bilayer}). This so-called Bernal stacking yields a unit cell of
four atoms(A1, B1, A2 and B2) resulting in four electronic bands. The
tight-binding Hamiltonian for bilayer graphene can be written
as\cite{Neto-2008,McCann-2006}
\begin{align}
\label{eq:11}
H_0 =& -t \sum_{\langle i,j\rangle,m,\sigma} \left(
  a^{\dagger}_{\sigma,mi}b_{\sigma,mj} + \text{h.c.} \right)\notag\\
&-t_{\perp} \sum_{\langle 1i,2j\rangle,\sigma} \left(
  a^{\dagger}_{\sigma,1i}b_{\sigma,2j} + \text{h.c.} \right),
\end{align}
where $t_{\perp}(\approx 0.35 \text{eV})$ is interlayer hopping through dimer
states, $m=1,2$ is plane index. In momentum space, the effective bilayer
Hamiltonian has the form of
\begin{align}
\label{eq:12}
\mathcal{H}_0 =
\begin{pmatrix}
0 & \phi^{*} \left( \mathbf{k} \right) & 0 & 0 \\
\phi \left( \mathbf{k} \right) & 0 & -t_{\perp} & 0 \\
0 & -t_{\perp} & 0 & \phi^{*} \left( \mathbf{k} \right) \\
0 & 0 & \phi \left( \mathbf{k} \right) & 0
\end{pmatrix},
\end{align}
The eigenstates of Eq.~(\ref{eq:12}) are four component spinors
${\psi_{\sigma} = \left( \psi_{\sigma\text{A1}}, \psi_{\sigma\text{B1}},
    \psi_{\sigma\text{A2}}, \psi_{\sigma\text{B2}}\right)^T}$,
where $\psi_{\text{A1,B1}}(\psi_{\text{A2,B2}})$ are the envelop functions
associated with the probability amplitudes at the respective sublattice sites of
the lower(upper) graphene sheet.

The Hamiltonian $\mathcal{H}_0$ determines the following spectrum of electrons
in a bilayer graphene. There are four valley-degenerate bands, $E^{\pm}_s$,
$s=\pm 1$, with\cite{Nilsson-2007}
\begin{align}
\label{eq:13}
E^{\pm}_s \left( \mathbf{k} \right) = \pm \left| \sqrt{\frac{t^2_{\perp}}{4} +
    \epsilon^2\left( \mathbf{k} \right)} + s\frac{t_{\perp}}{2} \right|,
\end{align}
where $\epsilon \left( \mathbf{k} \right)$ is the dispersion of a monolayer
graphene. The dispersion $E^{\pm}_{-1}$ describe low energy bands while
$E^{\pm}_{+1}$ describe higher energy bands as shown in Fig.~\ref{fig:bilayer}. One
can see that low energy excitations exhibit parabolic dispersion, while for
larger $k$ values, the linear $E-k$ behavior is recovered.

An important difference in the eigenfunctions between the monolayer and the
bilayer graphene is that in the latter case there are eight eigenstates (two
bands) for a given energy $E_i$ and fixed $k_y$. Accordingly, for four component
spinor $\psi_{\sigma}$, the pseudo-spin in bilayer graphene is defined as
eigenvalue of Dirac matrix
\begin{align}
\label{eq:14}
\gamma_1 = \begin{pmatrix}
0 & 0 & 0 & 1\\
0 & 0 & 1 & 0\\
0 & 1 & 0 & 0\\
1 & 0 & 0 & 0
\end{pmatrix}.
\end{align}
As a result, the pseudo-spin of state $E^{\pm}_s$ in bilayer graphene is $\mp
s$. So the lower band ($E^{+}_{-1}$) in Fig.~\ref{fig:bilayer} is associated
with pseudo-spin $+1$, and the upper band ($E^{+}_{+1}$) with $-1$. Therefore,
both inter- and intra-valley scattering do not flip the pseudo-spin. This
feature is totally different from monolayer structure. Moreover, in our
computation, the energy of incident wave satisfies $E_i \ll t_{\perp}$ as this
will likely be the experimental situation. So in region \Rmnum{1} and \Rmnum{5}
we have four of these states($\mathbf{k}_{1,2,3,4}$) correspond to propagating
waves($E^{+}_{-1}$) and the other four($\boldsymbol{\kappa}_{1,2,3,4}$) to
evanescent ones($E^{+}_{+1}$). These evanescent modes have an complex value of
the momentum in the $x-$direction and must be considered to fulfill the boundary
conditions. It is worth noting that the incident electron in bilayer lattice
could only reside on band $E^{+}_{-1}$ with pseudo-spin $+1$.

By solving Hamiltonian in Eq.~(\ref{eq:12}), one finds the eigenvectors($V$ is
external potential),
\begin{align}
\label{eq:15}
\psi^{\pm}_s \left( \mathbf{k} \right) = C
\begin{pmatrix}
\alpha \left( \mathbf{k} \right) \phi^{*} \left( \mathbf{k} \right)/ \left(
  E^{\pm}_s - V \right)\\
\alpha \left( \mathbf{k} \right)\\
\left( E^{\pm}_s -V \right) / \phi \left( \mathbf{k} \right)\\
1
\end{pmatrix}
e^{i \mathbf{k}\cdot \mathbf{r}},
\end{align}
where $C$ is normalization constant, and $\alpha \left( \mathbf{k} \right)$ is
defined as
\begin{align}
\alpha \left( \mathbf{k} \right) = - \left[ \left( E^{\pm}_s - V \right)^2 -
  \left| \phi \left( \mathbf{k} \right) \right|^2 \right] / t_{\perp} \phi \left( \mathbf{k} \right).
\end{align}
Note again that for propagating waves, $k_x$ is real; while for evanescent
solutions, $k_x$ is complex.

Similar to the case of the monolayer graphene, by calculating the group velocity
$v_x \left( \mathbf{k} \right) = \mathrm{d} \epsilon \left( \mathbf{k}
\right)/\hbar \mathrm{d} k_x$,
we can prove that $\mathbf{k}_{1,3}$ are right movers (transmission waves) while
$\mathbf{k}_{2,4}$ are left movers (reflective waves). On the other hand, to
select appropriate evanescent states, since $\kappa_{lx} (l=1,2,3,4)$ are
complex numbers, we should consider the asymptotic behavior of these states at
$\pm \infty$.

Finally, the general solutions for bilayer graphene can be expressed
as\cite{Bai-2007}(incident electron coming from $\mathbf{k}_1$)
\begin{subequations}
\begin{align}
\label{eq:16a}
&\Psi_{\text{\Rmnum{1}}} \left( \mathbf{r} \right) = \psi^{+}_{-1} \left( \mathbf{k}_1
\right) + r_1 \psi^{+}_{-1} \left( \mathbf{k}_2 \right) + r_2 \psi^{+}_{-1} \left( \mathbf{k}_4
\right)\notag\\
& \ \ \ \ \ \ \ \ \ \ \ \ + r_3 \psi^{+}_{+1} \left( \boldsymbol{\kappa}_2 \right) + r_4
\psi^{+}_{+1} \left( \boldsymbol{\kappa}_4 \right),\\
\label{eq:16b}
&\Psi_{\text{\Rmnum{3}}} \left( \mathbf{r} \right) = \sum^4_{l=1} \left[ f_l
  \psi^{-}_{-1} \left( \mathbf{q}_l \right) + g_l \psi^{-}_{+1} \left( \boldsymbol{\tau}_l \right)
  \right],\\
\label{eq:16c}
&\Psi_{\text{\Rmnum{5}}} \left( \mathbf{r} \right) = t_1 \psi^{+}_{-1} \left(
  \mathbf{k}_1 \right) + t_2 \psi^{+}_{-1} \left( \mathbf{k}_3 \right)\notag\\
& \ \ \ \ \ \ \ \ \ \ \ \ + t_3 \psi^{+}_{+1} \left( \boldsymbol{\kappa}_1 \right) + t_4
\psi^{+}_{+1} \left( \boldsymbol{\kappa}_3 \right).
\end{align}
\end{subequations}
Here $\mathbf{q}_l$ and $\boldsymbol{\tau}_l$ are the corresponding wavevectors
for propagating states ($E^{-}_{-1}$) and evanescent states ($E^{-}_{+1}$)
inside the barrier. Then by solving Eq.~(\ref{eq:4}) numerically in \Rmnum{2}
and \Rmnum{4}, and utilizing continuity of the wave functions on four
sublattices at boundaries, one may obtain the transmission coefficients for a
bilayer graphene lattice by using the same equations as in the case of monolayer
structure.

\section{Numerical results and discussion}
\label{sec:result}

\begin{figure}[htbp]

    \resizebox{80mm}{!}{\includegraphics[]{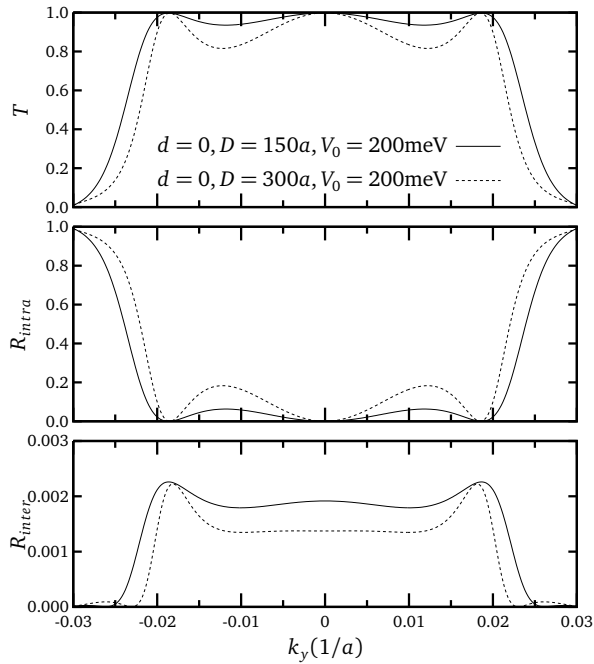}}

\caption[]{\label{fig:monoky} $k_y$ dependence of transmission coefficients in
  monolayer graphene for abrupt potential barrier($d=0$). The incident electron
  comes from valley $\mathbf{K}_{+}$ with energy $80$meV. The potential height
  is $200$meV with different widths $D=150a$ or $300a$. During the whole
  tunneling process, we always have $T+R_{intra}+R_{inter}=1$ to ensure current
  conservation. Note that different scaling are used in $R_{intra}$ and
  $R_{inter}$}

\end{figure}

\begin{figure}[htbp]

     \resizebox{80mm}{!}{\includegraphics[]{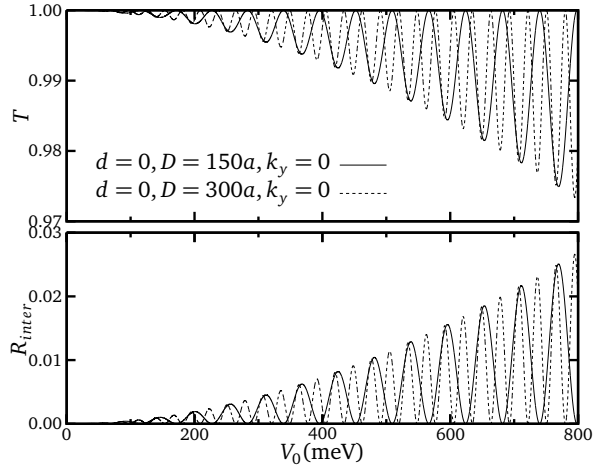}}

\caption[]{\label{fig:mononormallow} Transmission probability $T$ and
  inter-valley scattering amplitude $R_{inter}$ for normally incident electrons
  through monolayer graphene as a function of external potential height. The
  incident electron comes from valley $\mathbf{K}_{+}$ with energy $80$meV. The
  width of the barrier is $d=0$ and $D=150a$ or $300a$. For normal incident
  tunneling, we always have $R_{intra}\equiv 0$.}

\end{figure}

We first calculate the transmission probabilities of charge carriers through monolayer graphene lattice. The results are
depicted in Figs.~\ref{fig:monoky} and \ref{fig:mononormallow}. Fig.~\ref{fig:monoky} shows examples of $k_y$ dependence
of transmission probability for an abrupt potential barrier with height $V_0=200$meV. Under this potential barrier,
incident electron continues propagating as a hole in region \Rmnum{3}. The solid lines and dashed lines correspond to
the potential width $D=150a$ and $300a$, respectively. It is seen from the figure that, averagely speaking, the
intra-valley scattering coefficient is much larger than inter-valley scattering coefficient. This is because in
scattering process(see Fig.~\ref{fig:monolayer}), the intra-valley momentum transfer is much smaller than the
inter-valley momentum transfer($\sim 2 \left|\mathbf{K}_{+}\right|$), thus should have larger possibilities. It is also
clear from the figure that the barrier remains nearly perfectly transparent\cite{Katsnelson-2006,Bai-2007}($T\sim 1$)
for small $k_y$. This phenomenon is unique to relativistic quasiparticles that incident electrons can be scattered into
hole states inside the barrier. In this figure, the incident particle is in state $\mathbf{k}_1$ which comes from valley
$\mathbf{K}_{+}$. We can also choose the incident particle coming from valley $\mathbf{k}_3$. The difference of these
two results is very small, which means that valley discrepancy is very tiny in low-energy case.

Now let's examine the normally incident cases, i.e, $k_y=0$. The transmission coefficients are plotted in
Fig.~\ref{fig:mononormallow}, where we find that intra-valley scattering is totally suppressed. This can be understood
in terms of the conservation of pseudo-spin. In normal incident tunneling, pseudo-spin becomes a good quantum number
through the whole process. Therefore, spin flipping is not allowed when quasiparticles propagating in this system.
Hence, inter-valley scattering is the only reflection mechanism. However, since momentum transfer $\mathbf{k}_1 \to
\mathbf{k}_4$ is quite large, $R_{inter}$ is very small ($\sim 0.01$). So if we neglect the inter-valley scattering
probability\cite{Park-2008}, then monolayer graphene can be regarded as a condensed matter version Klein
paradox\cite{Katsnelson-2006,Bai-2007}. Furthermore, we can see that the transmission amplitude shows a resonant feature
that the envelope of $T$ deceases monotonically as $V_0$ increases. This can be understood as Dirac-like particle's
resonance tunneling, where the resonance frequency and magnitudes varies significantly with different barrier widths.
With increasing $D$, the resonance frequency increases obviously as shown in Fig.~\ref{fig:mononormallow}, while the
envelope of transmission amplitude remains the same. On the other hand, the magnitude of oscillation also depends
sensitively on barrier width $d$. By increasing $d$, inter-valley scattering amplitude falls noticeably. For example, in
case of $d=3a$, the inter-valley scattering amplitude falls to the order of $10^{-5}$ when $V_0=800$meV. Thus one can
achieve nearly perfect transmission even with very high barrier height by smoothing the potential step.

\begin{figure}[htbp]

 $\begin{array}{cc}
    \resizebox{80mm}{!}{\includegraphics[]{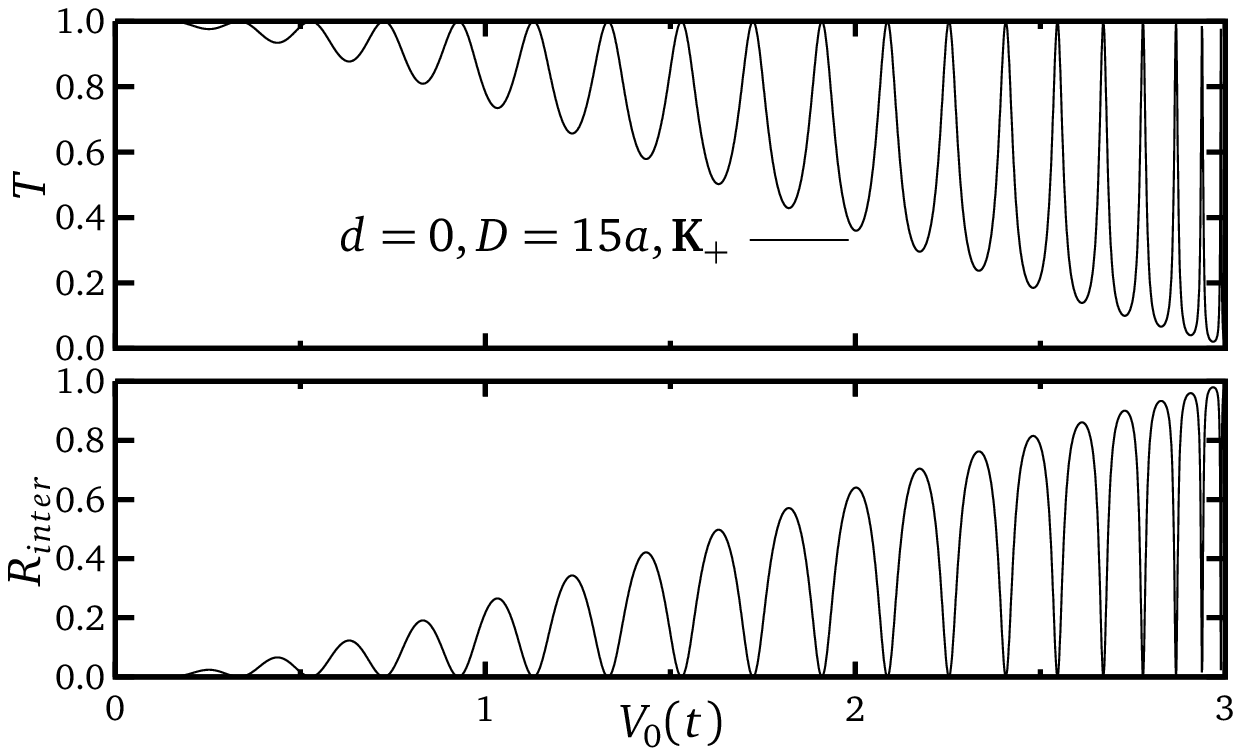}}\\
    \resizebox{80mm}{!}{\includegraphics[]{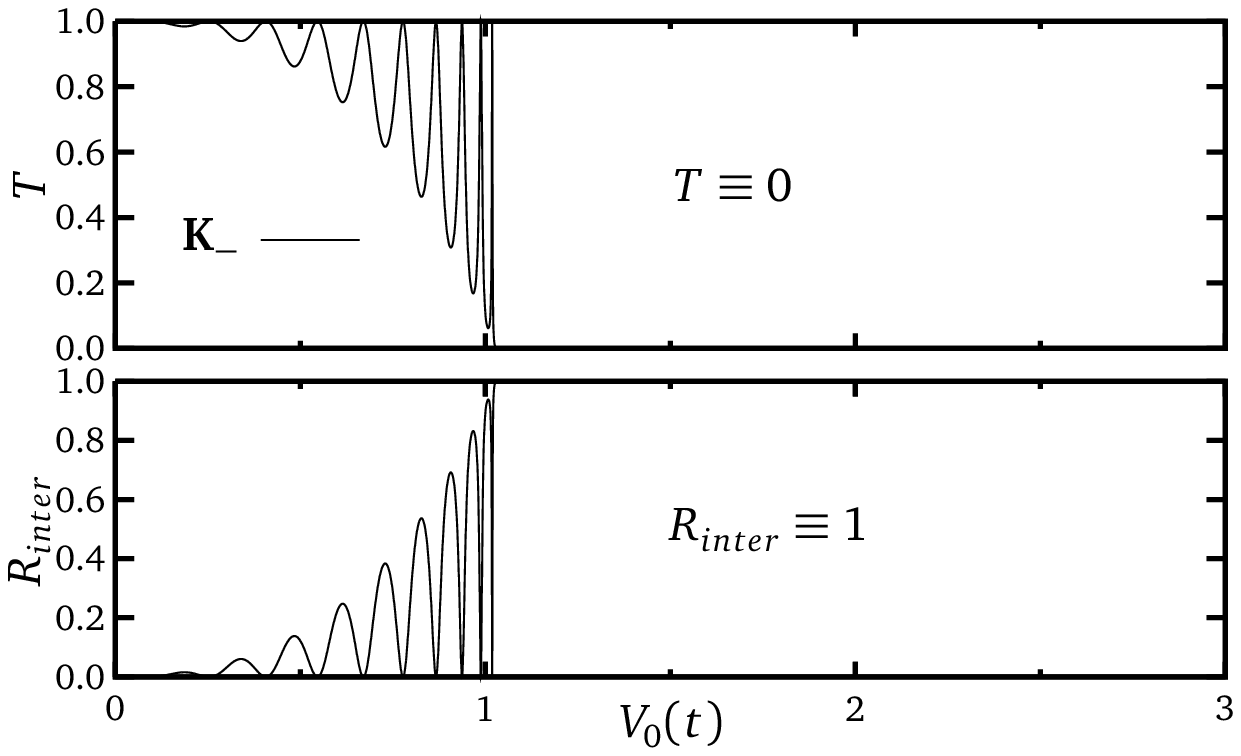}}
  \end{array}$

\caption[]{\label{fig:mononormalhigh} Transmission probability $T$ and
  inter-valley scattering amplitude $R_{inter}$ for normally incident electrons
  through monolayer graphene as a function of external potential height. The
  incident electron is of energy $80$meV and may have valley polarization of
  Dirac points $\mathbf{K}_{+}$ or $\mathbf{K}_{-}$. The potential width is
  $d=0$ and $D=15a$. Top: Incident electron comes from $\mathbf{K}_{+}$. Bottom:
  Incident electron comes from $\mathbf{K}_{-}$.}

\end{figure}

By now we have investigated low-barrier ($V_0 \ll t$) tunneling in monolayer
graphene by choosing incident particle as $\mathbf{k}_1$ which comes from
$\mathbf{K}_{+}$. If the incident electron comes from valley $\mathbf{K}_{-}$, our
result shows no obvious difference with valley $\mathbf{K}_{+}$. To amplify this
valley-contrasting feature, we have to look at the high-barrier ($V_0 \sim t
\approx 2.8$eV) limit of the tunneling problem. The results of normally incident
transmission are depicted in Fig.~\ref{fig:mononormalhigh}.

In this figure, one can see that the situation is completely different from
low-barrier scattering. Besides the resonant feature in the transmission
pattern, incident electrons coming from $\mathbf{K}_{+}$ and $\mathbf{K}_{-}$
have totally distinct tunneling properties. The origin of this valley
dissimilarity is due to the unique energy dispersion in graphene as shown in
Fig.~\ref{fig:monolayer}. At first, it is worth noting that pseudo-spin is a
good quantum number in normally incident tunneling and the incident electrons
might have two types of pseudo-spin, $+1 (\mathbf{K}_{+})$ or
$-1 (\mathbf{K}_{-})$. For low potential barrier, i.e. $V_0 < E_i + t$, there
are four propagating states in region \Rmnum{3}. Two of these states have
pseudo-spin $+1$, while the other two have pseudo-spin $-1$. Therefore, no
matter which incident state we select, the pseudo-spin inside the barrier can
always match the incoming quasiparticle. So in low-barrier limit, resonant
pattern is found in the transmission amplitude of both valleys.

On the other hand, for high potential barrier, i.e. $V_0 > E_i + t$, two
propagating states with pseudo-spin $-1$ in region \Rmnum{3} turn to evanescent
states\cite{Danneau-2008}, while the propagating states with pseudo-spin $+1$
remain the same. As a result, different incident electrons come from different
Dirac points $\mathbf{K}_{+}$ and $\mathbf{K}_{-}$ must have distinct
transmission properties. For incident electron in state $\mathbf{k}_1$ with
pseudo-spin $+1$, the potential barrier is always transparent. On the contrary,
for electron coming from state $\mathbf{k}_3$ with pseudo-spin $-1$, it cannot
penetrate the potential barrier if $V_0 > E_i + t$. Moreover, it is also shown
in Fig.~\ref{fig:mononormalhigh} that the possibility of inter-valley scattering
can be notably enhanced if we increase the potential height $V_0$. Using these
tunneling properties, one may obtain valley polarized electron
currents\cite{Rycerz-2007,Di-2007} in monolayer graphene by adding appropriate external gate
voltage.

Similar to low-barrier limit, we find that the magnitude of inter-valley
scattering amplitude also depends very sensitively on potential height $V_0$ and
the thickness of region \Rmnum{2}, \Rmnum{4} in Fig.~\ref{fig:potential}. If we
smooth the potential step by increasing $d$, then the envelope of inter-valley
scattering coefficient shrinks much smaller very quickly. Furthermore, although
inter-valley scattering requires a very large momentum transfer, we can still
achieve very high scattering probability by increasing the barrier height. Our
result shows that charge carriers in monolayer graphene, when subjected to a
high potential barrier, exhibit totally distinct tunneling features from Dirac
fermions.

\begin{figure}[htbp]

    \resizebox{80mm}{!}{\includegraphics[]{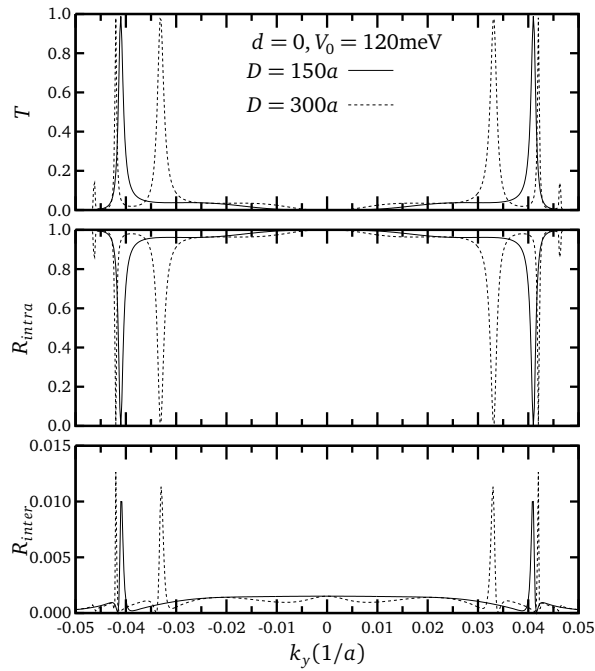}}

\caption[]{\label{fig:biky}$k_y$ dependence of transmission coefficients in
  bilayer graphene for abrupt potential barrier($d=0$). The incident electron
  comes from valley $\mathbf{K}_{+}$ with energy $80$meV. The potential height
  is $120$meV with different widths $D=150a$ or $300a$.}

\end{figure}

For the case of bilayer graphene, the computational results are shown in the
following figures. Some representatives $k_y$ dependence are depicted in
Fig.~\ref{fig:biky}, and these should be contrasted to the case of monolayer
in Fig.~\ref{fig:monoky}. The result is totally different since the bilayer
energy spectrum(see Fig.~\ref{fig:bilayer}) splits into two bands. One can see
that for potential barrier ($V_0$) higher than the energy of incident
quasiparticle ($E_i$), it remains nearly totally reflective for small $k_y$. In
normally incident ($k_y=0$) case, it becomes totally reflective even though
there are plenty of electronic states inside the
barrier\cite{Katsnelson-2006,Nilsson-2007}. This behavior, which is similar to
the tunneling property of non-chiral massive quasiparticles, is in obvious
contrast to monolayer graphene, where massless Dirac fermions are always
perfectly transmitted for small $k_y$\cite{Bai-2007}.

\begin{figure}[htbp]

$\begin{array}{cc}
    \resizebox{80mm}{!}{\includegraphics[]{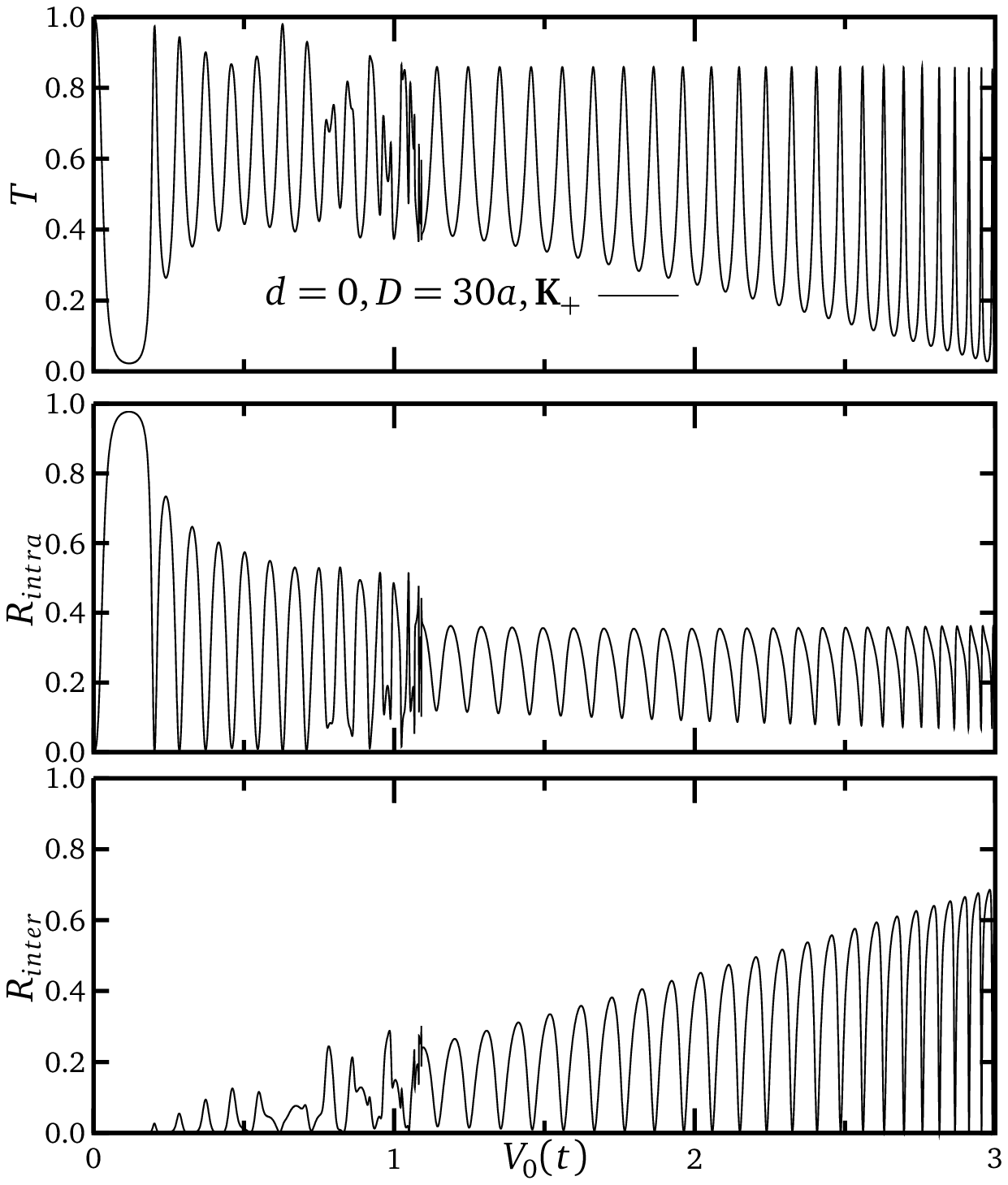}}\\
    \resizebox{80mm}{!}{\includegraphics[]{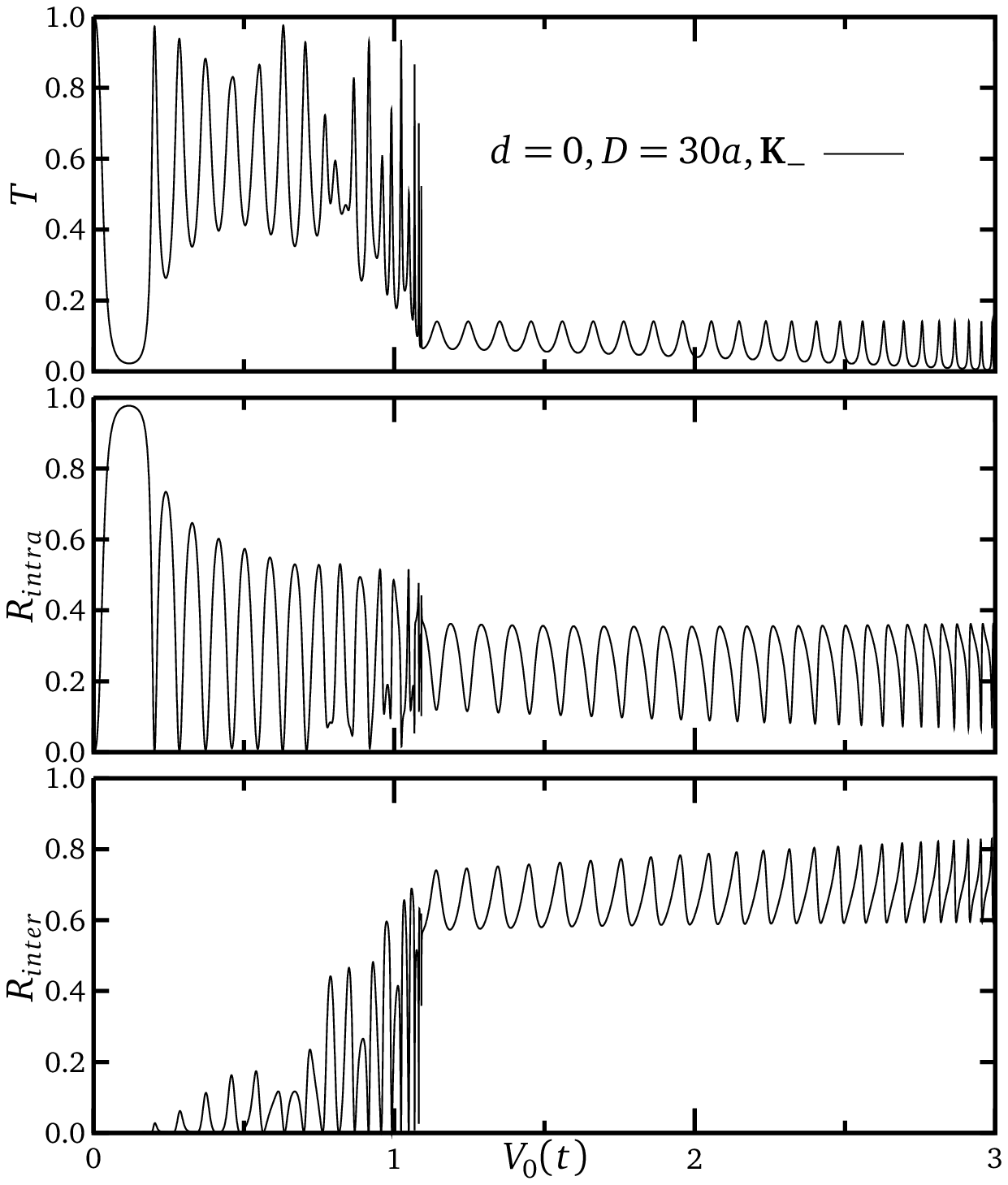}}
\end{array}$

\caption[]{\label{fig:binormal} Transmission coefficients for normally incident
  waves through bilayer graphene as a function of external potential height. The
  incident electron is of energy $80$meV and may have valley polarization of
  Dirac points $\mathbf{K}_{+}$ or $\mathbf{K}_{-}$. Potential width is $d=0$
  and $D=30a$. Top: Incident electron comes from $\mathbf{K}_{+}$. Bottom:
  Incident electron comes from $\mathbf{K}_{-}$.}

\end{figure}

To understand this feature, let's take a look at the result of normally incident
case as shown in Fig.~\ref{fig:binormal}. One can see in this figure that the
transmission probability $T\sim 0$ if
\begin{align}
\label{eq:17}
E_i < V_0 < E_i + t_{\perp},
\end{align}
and the transmission amplitudes exhibit nonresonant feature within this
range. This result is valley independent and can be explained in terms of
pseudo-spin conservation in normal incident tunneling. As one can see in
Fig.~\ref{fig:bilayer}, under the condition of Eq.~(\ref{eq:17}), all the
propagating states in region \Rmnum{3} are hole states $E^{-}_{-1}$ with
pseudo-spin $-1$. While the incoming wave, no matter which valley it comes from,
must have pseudo-spin $+1$. As a result, the only survival tunneling mechanism
is via evanescent hole states $E^{-}_{+1}$. Therefore, $\left( E_i,E_i+t_{\perp}
\right)$ becomes a non-penetrative region for bilayer graphene. This conclusion
is coincide with the $k_y$ dependence as we depicted in
Fig.~\ref{fig:biky}.

On the other hand, beyond this non-penetrative range, the resonant feature is
found again for all barrier heights as the pseudo-spin of propagating states in
region \Rmnum{3} can always match the incoming electron. Further more, valley
difference is not obvious in low energy limit when $V\ll t$, which is the same
as the case in monolayer graphene under low-barrier limit. While for high
barrier tunneling, the transmission patterns of different Dirac points are
notably distinct. This distinction is due to the inter-valley momentum transfer
$\left| \mathbf{k}_1 - \mathbf{k}_4 \right| = 2|\mathbf{k}_1|$
is larger than
$\left| \mathbf{k}_3 - \mathbf{k}_2 \right| = 2|\mathbf{k}_3|$
So intrinsically the probability of inter-valley scattering amplitudes
$\mathbf{k}_1\to \mathbf{k}_4$ should smaller than $\mathbf{k}_3\to \mathbf{k}_2$.
In Fig.~\ref{fig:binormal}, this difference is amplified by high potential
barrier and shows the existence of valley polarized current in bilayer
graphene\cite{Rycerz-2007}. Moreover, in Fig.~\ref{fig:binormal}, the patterns
of reflection coefficient $R_{intra}$ are almost the same. This can be
understood as intra-valley momentum transfer in two valleys are equal to each
other
\begin{align}
\label{eq:18}
\left| \mathbf{k}_1 - \mathbf{k}_2 \right| = \left| \mathbf{k}_3 - \mathbf{k}_4
\right|.
\end{align}

Finally, we also point out that the dependence of transmission amplitude on $d$ and $D$
in bilayer graphene is the same as the result in monolayer structure. If we
increase the width of region \Rmnum{3}, the envelope of $T$ curve does not
change while the period of resonant oscillation decreases. On the other hand, if
we smooth the potential step by broadening the width of region \Rmnum{2} and
\Rmnum{4}, the amplitude of inter-valley scattering falls very quickly to
smaller than $10^{-5}$. However, intra-valley scattering probability does not
depend on $d$ in our computation. Thus we can tune the proportion of these two
scattering mechanisms separately by adding arbitrary electronic fields at the
edges of the external potential barrier.

Before closing, just a comment on our calculation. Actually, previous studies of tunneling properties in monolayer
(bilayer) graphene that were based on a continuum model have used a massless Dirac fermion (massive excitations)
approximation, which is expected to be accurate for low-energy
quasiparticles\cite{Novoselov-2005,McCann-2006,Katsnelson-2006,Bai-2007,Nilsson-2007,Neto-2008}. Hence, inter-valley
scattering is rarely considered in investigating the transport properties of graphene systems\cite{Morpurgo-2006}
because of the large separation of Dirac points in momentum space. Moreover, valley contrasting physics is also absent
in low-energy quasiparticle's tunneling features in this system. However, our result shows that, although low-energy
tunneling in monolayer (bilayer) graphene is sufficiently described by massless (massive) fermion approximation, in
high-energy limit there will be a significant deviation between this approximation and the real
system\cite{Plochocka-2008}. With the increasing of barrier height, inter-valley scattering can be realized and
amplified significantly as shown in Figs.~\ref{fig:mononormalhigh} and \ref{fig:binormal}. At the same time, we find
that valley discrepancy in the tunneling problem would be gradually magnified by strengthening the external potential.
These effects, which are distinct from the properties of low-energy limit, lead to the possibility of generating,
detecting and controlling valley polarized currents in graphene systems by electric means\cite{Rycerz-2007,Di-2007}.

\section{Summary and conclusion}
\label{sec:summary}

Based on a tight-binding model, we have investigated single particle tunneling
properties through potential barrier in monolayer and bilayer graphene
lattices. The normal incidence and angular dependent transmission probabilities
for two kinds of graphene structure have been numerically calculated. Our result
illustrates that the tunneling behavior in graphene is much richer than what was
anticipated in massless Dirac fermion approximation. Furthermore, it is shown
that both of the intra- and inter-valley scattering probability are strongly
dependent on the applied external potential, forming the basis for the
valley-based electronics applications of these systems. Our
results may provide an important reference to the design of electron
devices based on graphene materials.

\end{document}